\documentstyle[11pt,newpasp,twoside,graphicx,natbib]{article}

\def\edcomment#1{\iffalse\marginpar{\raggedright\sl#1\/}\else\relax\fi}
\reversemarginpar

\def\percmsq{{$\rm cm^{-2}$}}
\newcommand{\chandra}{\textit{ Chandra}}
\newcommand{\rxte}{\textit{ RXTE}}
\newcommand{\onee}	{\mbox{\rm\,1E~1740.7--2942}}
\newcommand{\Grs}	{\mbox{\rm\,GRS~1758--258}}
\newcommand{\cyg}	{Cyg~X-1}
\newcommand{\lumin}	{\mbox{$\rm\,ergs\,s^{-1}$}}
\newcommand{\xray}	{\mbox{X-ray}}
\newcommand{\eflux}	{\mbox{$\rm\,ergs~cm^{-2}~s^{-1}$}}
\newcommand{\asec}	{\mbox{$^{\prime \prime}$}}
\newcommand{\aprx}	{\mbox{$\sim$}}
\newcommand{\ltsim}{\lower.5ex\hbox{$\; \buildrel < \over \sim \;$}}

\begin{document}

\title{The Spectrum and Accurate Location of GRS~1758-258}
\author{William A. Heindl}
\affil{Center for Astrophysics and Space Sciences, University of
California, San Diego, La Jolla, CA 92093, U.S.A.}

\author{David M. Smith}
\affil{Space Sciences Laboratory, University of California, Berkeley, 
Berkeley, CA  94720, U.S.A.}

\begin{abstract}
We observed the ``micro-quasar'' \Grs\
four times with \chandra.  Two HRC-I observations were made in
2000 September-October spanning an intermediate-to-hard spectral
transition (identified with \rxte).  Another HRC-I and an ACIS/HETGS
observation were made in 2001 March following a hard-to-soft
transition to a very low flux state.  The accurate position (J2000) of the
\xray\ source is RA $=$ 18 01 12.40,
Dec $=$ $-$25 44 36.0 (90\% confidence radius $=$ 0\asec.6),
consistent with the purported variable radio counterpart.  All images
are consistent with \Grs\ being a point source, indicating that any
bright jet is less than \aprx1\,light-month in projected length,
assuming a distance of 8.5\,kpc. The March spectrum is well-fit with a
multi-color disk-blackbody with an inner temperature of $0.50 \pm
0.01$\,keV, interstellar absorption of $n_H = (1.59 \pm 0.05)\times
10^{22}$\,\percmsq, and (un-absorbed) 1$-$10\,keV luminosity of $\rm
4.5 \times 10^{36} (D/8.5\,kpc)^2$\,\lumin.  No narrow emission lines
are apparent in the spectrum and upper limits to line equivalent
widths are given.
\end{abstract}

\section{Introduction}
      
\Grs\ and its sister source, \onee, were the first objects dubbed
``micro-quasars''.  Their \xray\ spectra are typical of Galactic black
hole candidates (BHCs), and they are associated with time variable
cores of double-lobed radio sources, reminiscent of extra-Galactic
radio sources.  This morphology, seen on a parsec scale within the
Milky Way, earned them their nickname. \Grs\ and \onee\ are the
brightest persistent sources in the Galactic bulge above $\sim$50~keV
\citep{Su91}.  Their timing characteristics are typical of the black
hole low/hard state \citep{Mai99,Smi97,Hei93,Su91}, and they
consistently emit near their brightest observed levels, although
they vary over times of days to years.  Their \xray\ emission
properties are readily likened to the canonical BHC, \cyg.  In fact,
together with \cyg, they are the only known persistent, low-state
BHCs, and all three sources have maximum luminosities around $3 \times
10^{37}$ergs s$^{-1}$. Radio jets have now been observed in
\cyg, furthering the similarity \citep{Fen00}.

\Grs\ and \onee\ are, however, quite different from the Galactic {\em
superluminal} radio sources more typically thought of as
micro-quasars: GRS~1915+105 and GRO~J1655-40.  The \xray\ emission
from these objects is much brighter and more spectacularly variable.
Their radio jets, too, are much brighter and are highly variable,
being unresolved or absent except during exceptional ejection events
which last only weeks. In contrast, the radio lobes of  \Grs\ and \onee\
are quite stable \citep{Mir99}.

\begin{figure}
\label{f_lc}
\centerline{\includegraphics[width=4.5in,bb=1 18 434 283,clip]{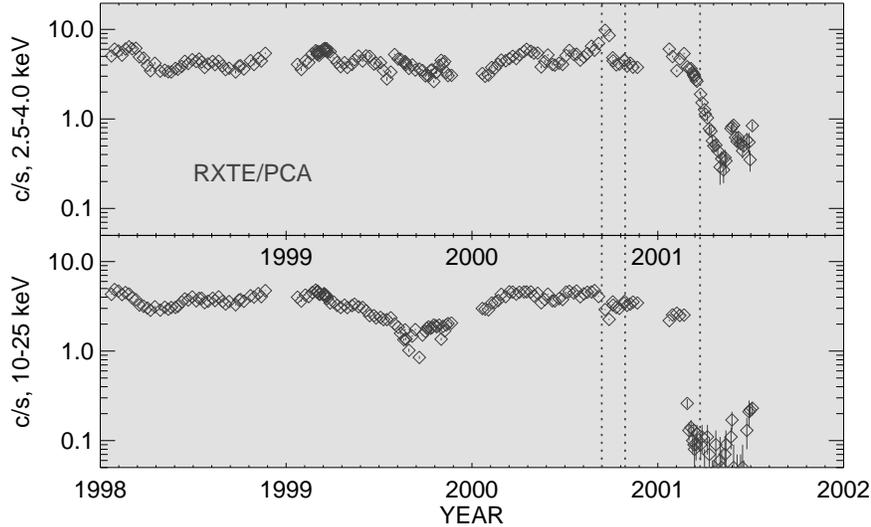}}
\caption{The \rxte/PCA light curve of \Grs\ in two energy bands.
Our \chandra\ observations (see Table~1) are indicated by dashed
vertical lines.  Observations 400163 and 400164 were made consecutively
and so appear as a single line near 2001.2. The 1996-1997 flux history
appears very similar to 1998 with the source remaining quite stable
within a factor of \aprx2.}
\end{figure}

\begin{table}
\caption{\label{t_obs}Observations}
\begin{center}
\begin{minipage}{0.7\textwidth}
\begin{tabular}{lcccc} \hline\hline
Seq. \# & Date		& Inst.	& Exp.	& Rate \\
	& 		&  		&(ksec)	& (counts/s)\\ \hline
400085	& 2000 Sep 11.2	&  HRC-I	& 1	&  11.4\\
400131	& 2000 Oct 27.4	&  HRC-I	& 10	&  4.2 \\
400164	& 2001 Mar 24.3	&  HRC-I  	& 10	&  7.8 \\ 
400163	& 2001 Mar 24.4	&  ACIS-S	&	& \\
	&		&  /HETGS	& 30	&  6.8\footnote{order
0 rate, significantly piled up}\\\hline\hline
\end{tabular}
\end{minipage}
\end{center}
\end{table}

\section{Observations}
We observed \Grs\ four times with \chandra.  Table~1 lists the
observation dates and durations. Two HRC-I observations were made in 2000
September-October spanning an intermediate-to-hard spectral transition
\citep[identified with \rxte\ ][]{Smi01b,Smi01a}.  
Another HRC-I and an ACIS/HETGS observation were made back-to-back in
2001 March following a dramatic hard-to-soft transition to a
\emph{very} low flux state.  Figure~\ref{f_lc} shows the \rxte/PCA
light curve with our \chandra\ observations indicated.

\section{Imaging and the Accurate Position}
\begin{figure}
\label{f_image}
\vspace{2ex}
\centerline{\includegraphics[width=3.in,bb=-20 -20 423 423]{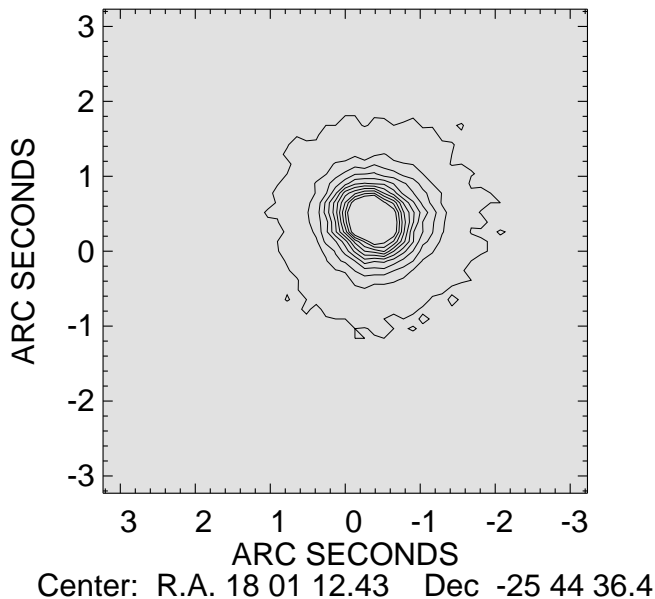}}
\caption{HRC-I image from observation 400131 centered on the indicated
pointing direction.  \Grs\ appears point-like, with no indication of
\xray\ jets.}
\vspace{1.ex}
\end{figure}

Figure~\ref{f_image} shows the HRC-I image from observation 400131.
In this and the other HRC observations, the \Grs\ image is consistent
with the HRC point spread function, indicating that the source is
point-like at the sub-arcsecond level.  Assuming a distance of
8.5\,kpc, this says that no strong jets are present on a physical
scale of a light-month, the presence of arcminute scale radio jets
notwithstanding.  This is perhaps not surprising, as the timescale
associated with producing the parsec-sized radio lobes would be years
rather than months, and in fact the radio lobes are observed to be
quite stable over years\citep{Mir99}.  Thus, while de facto not ruled
out, it is not required that a \emph{persistent} small-scale \xray\
jet be present to produce the observed radio lobes.

\begin{figure}
\label{f_posn}
\centerline{\includegraphics[width=3.in, bb=-20 -20 423 423]{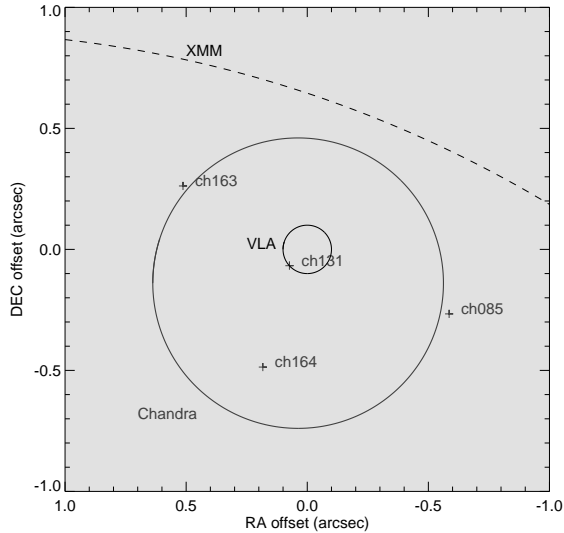}}
\caption{Error circles for \Grs. The 90\% confidence \chandra\ error
circle (0\asec.6 radius) includes the 0\asec.1\ VLA
radio position of (source ``VLA-C'')\citep{Mar98}. Coordinates are
offsets from the radio position: (J2000) RA $=$ 18 01 12.395, Dec $=$
-25 44 35.90.  The recent error circle from \emph{XMM} is also
indicated \citep{Gol01}.}
\vspace{1.ex}
\end{figure}

Figure~\ref{f_posn} shows the best fit source locations from our
four \chandra\ observations as well as the estimated 90\% confidence
region derived from the average of the four positions.   The
ACIS-S/HETGS position (400163=ch163) was based on the zeroth-order
image.  While the image was significantly piled up, we
expect a negligible effect on the position.  The
\chandra\ error circle is centered at (J2000) \emph{\bf{ RA $=$ 18 01
12.40, Dec $=$ $-$25 44 36.0}} and has an estimated 90\%
confidence radius of 0\asec.6 based on \chandra's absolute aspect
accuracy and averaging of four independent observations.  The
coincidence of the sub-arcsecond VLA and \chandra\ error circles seals
the association of \Grs, the \xray\ source, with the variable radio source
\citep[``VLA-C'', ][]{Mar98}.

\section{Spectrum}
\begin{figure}
\label{f_spec}
\centerline{\includegraphics[height=2in, bb= 50 61 423 423, clip]{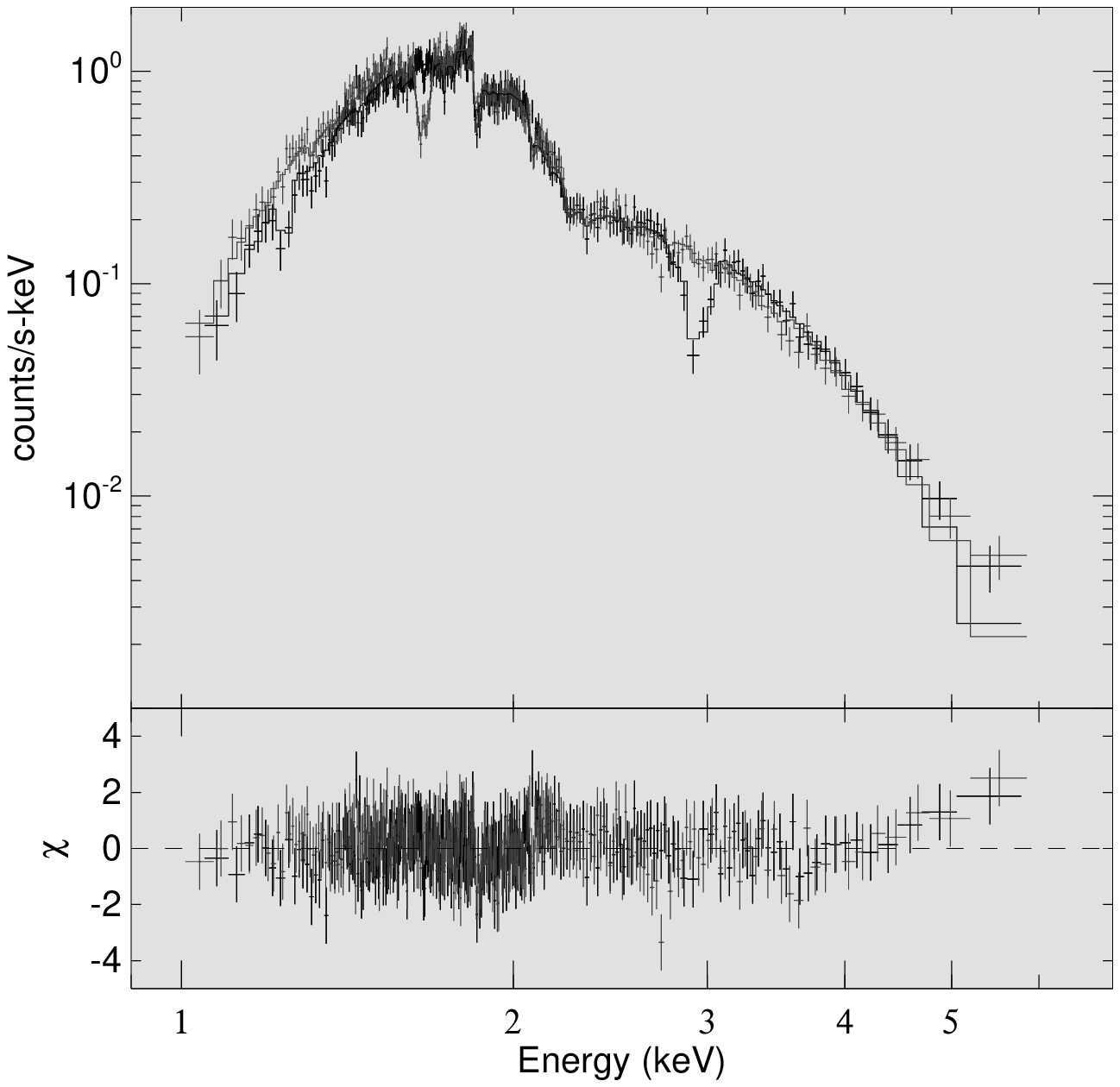} 
	\hspace{0.25in} \includegraphics[height=2in, bb=3 14 423 423]{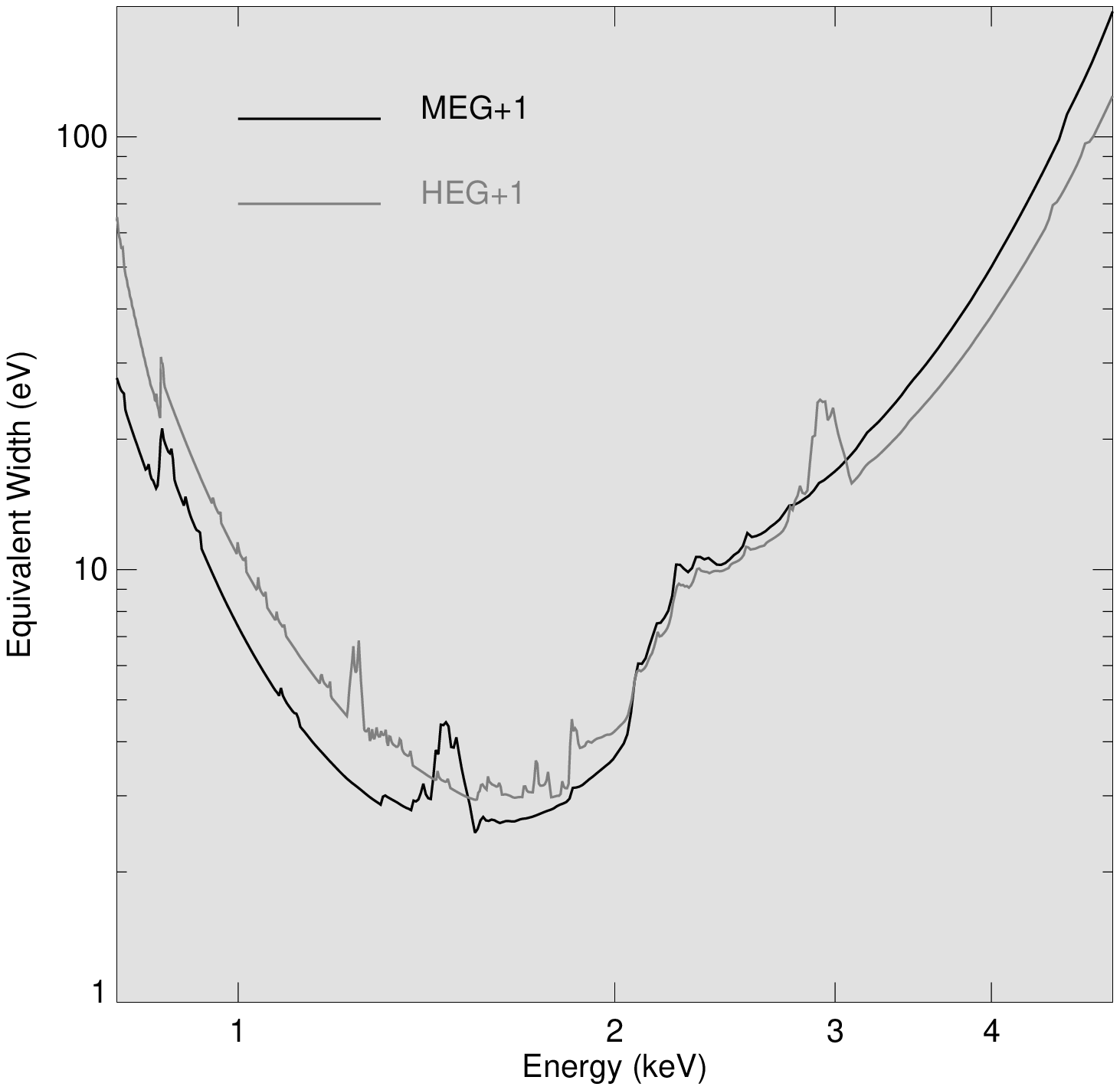}}
\caption{Left: The HEG first order spectra of \Grs\ fit with a multi-color
disk-blackbody spectrum and interstellar absorption. Right: The MEG
and HEG order +1 sensitivity (5$\sigma$ to narrow lines in the \Grs\ spectrum.}
\end{figure}

To reduce pile-up in the ACIS/HETGS observation, we used a 1/2 chip
sub-array with a 1.7\,s frame time.  However, the counting rates in
the MEG first order spectra were still about 2\,counts/s, resulting in
mild (\ltsim10\%) pileup around 2\,keV.  For this presentation, we
therefore restrict spectral fitting to the first order HEG spectra
which did not suffer from pile-up.  Figure~\ref{f_spec} shows the HEG
Order $\pm 1$ spectra rebinned to a minimum of 100 counts per bin.  The
data are well described by a multi-color disk-blackbody model and
interstellar absorption (XSPEC: ``phabs*diskbb'').  The positive
deviations above 4\,keV are indicative of a weak power-law component
(\aprx 2\% of the 1$-$10\,keV unabsorbed flux) seen in joint fits with
a contemporaneous \rxte/PCA (3$-$15\,keV) observation.  Table~2 lists
the best fit parameters to the HEG spectra alone.

Preliminary inspection of both the HEG and MEG spectra showed no
strong emission lines. Figure~\ref{f_spec} shows the 5$\sigma$
sensitivity to narrow emission lines. Because the \Grs\ spectrum is so
soft, the line sensitivity is a strong function of energy, varying by
nearly two orders of magnitude between 1$-$5\,keV.  The sensitivity was
based on the best fit spectrum (see above) and the effective area of
the ACIS/HETGS combination.


\begin{table}
\caption{\label{t_fit}Best fit absorbed disk-blackbody spectrum.}
\begin{center}
\begin{minipage}{0.8\textwidth}
\begin{tabular}{lcccc} 
\hline\hline
$\rm n_H$ ($\rm \times10^{21}$\,\percmsq) &	$\rm 15.9 \pm 0.5$ \\
$\rm kT_{in}$~(eV)	& $\rm 505 \pm 7$ \\
Flux~(1--10\,keV, \eflux)	& $1.8 \times 10^{-10}$ \\ 
$\chi^2_{red}$/dof & 0.66/544 \\ \hline\hline
\end{tabular}
\end{minipage}
\end{center}
\end{table}

\section{Discussion}

During more than 5 years' monitoring with the \rxte\ prior to 2001
March, the \Grs\ hard \xray\ spectrum was always dominated by a hard
power law with photon index $\Gamma \sim 1.5 - 2.5$ \citep{Smi01b}
with occasional appearance of a weak thermal component
\citep{Mer94,Hei98,Lin00}.  As shown in Figure~\ref{f_lc}, \Grs\ made an
abrupt state change in 2001 March.  The hard flux dropped by an order
of magnitude in a few days, leaving the thermal component seen in
Figure~\ref{f_spec}.  Based on relative luminosity, however, the current
soft state is not a \emph{high}/soft state.  Rather it is
significantly less luminous than the low/hard state in this source.
This can be contrasted to Cyg~X-1 and the soft \xray\ transients,
where the \emph{high}/soft state is more luminous.  Rather, this seems
to be a low-luminosity state which is fading into quiescence
(Figure~\ref{f_lc}).  Finally, we note that the measured column
density is consistent with previous measurements \citep{Mer94,Lin00,Gol01} 

Since strong jet ejections are generally associated with the ``very
high state'' and transitions from the ``off'' to high/soft states in
\xray\ transients \citep{Fen01}, it is perhaps not surprising that no
jet emission appeared in our low/hard state observations (Sep-Oct
2000) and the recent transition observation (Mar 2001).  Perhaps our
best opportunity will come when (if?) \Grs\ makes a transition once
again to its normal, low/hard state.  We have an approved \chandra\
cycle 3 proposal to monitor the morphology of \Grs\ and hope to
observe a jet ejection.\\


\end{document}